\documentstyle[11pt]{nature} \title{
How did the metals in a giant star originate?} 
\mainauthor{} \author{P. Bonifacio \affiliation { Istituto Nazionale
di Astrofisica -- Osservatorio Astronomico di Trieste, Via Tiepolo 11
34131, Trieste, Italy}, M. Limongi \affiliation{Istituto Nazionale di
Astrofisica -- Osservatorio Astronomico di Roma, Via Frascati 33, 00040
Monteporzio Catone, Roma, Italy}, A. Chieffi \affiliation{ Istituto di
Astrofisica Spaziale, CNR, via Fosso del Cavaliere,00133 Roma, Italy}}

\begin{document} \summary{~} \maketitle

\raisebox{13cm}[-13cm]{\emph{Nature} {\bf 422} (2003),	(issue 24 April)}

The chemical composition of stars with
extremely low metal contents
(taking ``metals'' to mean all elements other
than hydrogen and helium)
provides us with information on the masses of the stars
that produced
the first metals. Such a direct connection is not possible, however,
if the
surface  of  the	star  has  been  polluted  by enriched material,
either dredged from the
star's interior or transferred from a companion
star. Here we argue that, in the case
of HE0107-5240 (ref. 1), the most
iron
poor star known, the oxygen abundance
could	be  a  discriminant:  a
ratio  of  [O/Fe] exceeding +3.5 would favour a pristine origin of
metals, whereas an [O/Fe] ratio of less than  +3 would favour the
pollution
hypothesis. Using this criterion, we suggest how the required
information on oxygen
abundance might be obtained.

HE0107-5240 shows carbon, nitrogen
and sodium enhanced, respectively,
by factors of $10^ 4$,  $10^{2.3}$
and 10 relative to iron,whereas magnesium,
which is usually more abundant in metal-poor stars, is present in
almost
the same amounts as iron; no single supernova (SNII) model seems to
show this
pattern\cite{cl2002}. Although several models may be worth considering,
we evaluate two here: a
pristine origin due to the combined enrichment
of at least two SNIIs, or a pollution of
the surface of the star after
its formation.

If the metals are pristine, then
HE0107-5240 must have formed from a cloud
enriched by the ejecta of at least two
zero-metallicity SNIIs of quite
different initial masses. The first supernova, which was
presumably rather
massive, would have produced the light elements observed in
HE0107-5240,
but none of the heavier ones, because of an extensive fallback, up to
the
base  of  the  helium  shell.  This  would have prevented the ratios
[C/N] and[Mg/C] from becoming too high. The second supernova, which
was less massive,would have provided all of the elements heavier than
and including magnesium.

In the case of pollution enrichment, on the contrary, 
HE0107 -5240 would be
either a low-mass, zero-metallicity star that accreted
matter enriched
by the first generation of SNII, or a second-generation low-mass star
of
very low metallicity. 
The high [C,N,Na/Fe] ratios would be the result
of
subsequent enrichment of the surface of the star, due either to an
internal process or to
the  accretion  of  matter  synthesized  by  an 
Asymptotic Giant Branch (AGB)
star in a binary system.

One way to discriminate between pristine 
and pollution origins of C and N is to
measure the abundance of
oxygen. In the pristine case, the more massive supernova
would provide a
large amount of oxygen, leading to a high [O/Fe] ($> +3.5$),
as can
be derived from the yields of zero-metal massive stars\cite{cl2002}. 
In the
pollution case, a lower [O/Fe] is expected 
($< +3.0$), as implied by the yields of zero-metal, intermediate-mass
stars\cite{clds}. An upper limit of 
0.1 pm (where 1 pm is $10^{-12}$ m) on the
equivalent width of the [OI] 630-nm line would provide a limit on
oxygen, namely [O/H] $< +2.3$ or [O/Fe] $< +3$. 
With the Ultraviolet Visual Echelle Spectrograph on the 8.2m
                  VLT telescope, we estimate that such a limit
could be achieved in about 30 h of exposure. 
The situation is slightly
better for OH ultraviolet lines, for which we estimate that
an upper
limit of [O/Fe] $<+2.0$ could be derived even at moderate signal-to-noise

ratios ($> 20$), which should be achieved in
about 4 h of exposure. At
this signal-to-noise  ratio,  discrimination  between
[O/Fe]$<+ 3.0$ and [O/Fe]$>+ 3.5$ is possible; however,
lower oxygen abundances
would be very difficult to measure, whatever the signal-to-noise
ratio. As high-resolution spectra have already been obtained,including
the ultraviolet region containing the OH lines, for a total of about 
30 h of exposure (ESO/ST-ECF Science Archive Facility), it may
now be possible to
carry out the test that we propose.

Among extremely metal-poor stars, oxygen has been observed only in CS
22949-037 (ref. 4), for which [O/Fe]=+2.0; 
all other known stars have
oxygen features that are
below the detection
threshold. However, CS 29498-043 (ref. 5)
has a very similar pattern
of marked overabundance of the lighter elements; its oxygen abundance,
which has not yet been measured, may prove to be strongly
enhanced as well.

We note that the high [C/N] 
observed in HE0107-5240
is difficult to explain  in  terms  of	pollution,  whether
internal or
from an AGB companion, because of the high abundance of primary
nitrogen
in both cases. We argue that a pristine origin is more likely, which
implies that
the [O/Fe] ratio is higher than in any other known metal-poor
star.


\begin{thebibliography}{10}


\bibitem[<1>]{chris} 
Christlieb, N. et al. Nature 419, 904-906 (2002). 
\bibitem[<2>]{cl2002} Chieffi, A. i\&
Limongi, M. Astrophys. J. 577, 281-294 (2002). 
\bibitem[<3>]{clds} Chieffi, A., Limongi,
M., Dom\`\i nguez, I. \& Straniero, O. in New
Quests in Stellar Astrophysics: The Link Between Stars and Cosmology
(eds Chavez, M., Bressan, A., Buzzoni, A. \& Divakara, M.) 47-54
(Kluwer Academic, Dordrecht, The Netherlands, 2002). 
\bibitem[<4>]{dep} Depagne,
E. et. al. Astron. Astrophys. 390, 187-198 (2002). 
\bibitem[<5>]{ao} Aoki, W., Norris,
J. E., Ryan, S. G., Beers, T. C. \& Ando, H.  
Astrophys. J. 576, L141-L144 (2002).

\end{thebibliography}
\end{document}